# Design of a robust superhydrophobic surface: thermodynamic and kinetic analysis.

Anjishnu Sarkar[1], Anne-Marie Kietzig[*1]
[1]Department of Chemical Engineering, McGill University, Montreal H3A0C5, QC, Canada
[*]Corresponding author. Email address : anne.kietzig@mcgill.ca

The design of a robust superhydrophobic surface is a widely pursued topic. While many investigations are limited to applications with high impact velocities (for raindrops of the order of a few m/s), the essence of robustness is yet to be analyzed for applications involving quasi-static liquid transfer. To achieve robustness with high impact velocities, the surface parameters (geometrical details, chemistry) have to be selected from a narrow range of permissible values, which often entail additional manufacturing costs. From the dual perspectives of thermodynamics and mechanics, we analyze the significance of robustness for quasi-static drop impact, and present the range of permissible surface characteristics. For surfaces with a Young's contact angle greater than 90° and square micropillar geometry, we show that robustness can be enforced when an intermediate wetting state (sagged state) impedes transition to a wetted state (Wenzel state). From the standpoint of mechanics, we use available scientific data to prove that a surface with any topology must withstand a pressure of 117 Pa to be robust. Finally, permissible values of surface characteristics are determined, which ensure robustness with thermodynamics (formation of sagged state) and mechanics (withstanding 117 Pa).

## 1. Robust superhydrophobic wetting states: surface characteristics

### 1.1. Introduction: Origin of penetration depth

As a droplet settles on a surface, three interfaces are formed, namely solid-liquid (SL), liquid-air (LA) and solid-air (SA). The mutual orientations of the interfaces determine the area occupied by each interface, also known as interfacial area. Based on the magnitudes of the interfacial areas, surface wetting can be broadly classified into two regimes, homogeneous and heterogeneous. A homogeneous wetting regime is marked by complete penetration of liquid inside the roughness valleys, and consequently, a lack of a liquid-air interface under the droplet. The apparent contact angle (APCA) for the homogeneous wetting regime is determined by the Wenzel equation [1]. A heterogeneous wetting regime is characterized by a composite liquid-air interface under the drop. A heterogeneous wetting regime with no liquid penetration is characterized by the Cassie equation [2]. A homogeneous regime and a heterogeneous regime with no penetration are also termed as Wenzel wetting state and Cassie wetting state, respectively. Current literature suggests the existence of a heterogeneous wetting regime with partial liquid penetration, also termed as metastable Cassie states [3-10]. A metastable Cassie state is characterized by its penetration depth, i.e. the degree of liquid penetration inside the roughness valleys and the geometric configuration of LA interface. Distinct metastable Cassie states, corresponding to unique values of penetration depth and/or interfacial orientation have been experimentally confirmed using various imaging and acoustic techniques [7, 8, 10].

Recently, we have deduced a characteristic set of equations which provides an implicit correlation of penetration depth of a liquid with the apparent contact angle [11]. Penetration depth depends on the manner, in which a surface and drop come in mutual contact [12, 13]. Many of the published results on wetting experiments involve deposition of water droplets from the top [13-18]. Here, the LA interface comes into contact with the apex of the surface roughness (Cassie state). Drop deposition from the top can be quasi-static or velocity driven. A quasi-static deposition onto the surface is characterized by a virtually stationary drop dispense at contact. The velocity driven deposition involves forcible impingement of a drop onto a surface. Both the aforementioned cases are known to cause an irreversible wetting transition from the Cassie to the Wenzel state [18-22]. If the pressure imparted by the drop on the surface (wetting pressure) exceeds the surface energy required in penetrating a unit volume of the roughness valleys (antiwetting pressure), a wetting transition can be observed. Superhydrophobic robustness is exhibited by a robust heterogeneous wetting regime, and is in direct correlation with the mode of drop deposition (velocity driven/quasi-static). Research on velocity driven deposition is focussed on the investigation of robustness of a superhydrophobic state. A surface is considered to be robust if it withstands the impact of a falling raindrop, with a typical terminal velocity of a few meters per second [23-26]. Although quasi-static drop transfer is relevant with several applications, current literature lacks the necessary surface characteristics that lead to robustness [27-31]. As the quasi-static mode of drop deposition allows formation of a free energy minimized wetting state, the robustness of the heterogeneous wetting regime can be investigated from the dual perspectives of thermodynamics and mechanics. Static contact angle measurement (CAM) is the most common case where a wetting state is formed as a result of quasi-static drop dispense. It is possible that the drop is accidentally dispensed at a sub-millimeter height above the substrate. Without considering the contribution of external factors such as steady drop dispense or vibrations, and within the tenets of gravity driven kinematics, the height of drop dispense is translated to an impact velocity at impact. We postulate that the maximum possible margin of error encountered with CAM is a 0.5 mm of accidental height, which corresponds to an impact velocity of 100 mm/s. Since the 100 mm/s margin happens to be the minimum impact velocity for drop impingement on a surface, drop impact is virtually indistinguishable from quasi-static liquid dispense on a surface [32-34]. Robustness, in this case, reflects the ability of the surface to withstand impact velocities less than 100 mm/s. Thus, the domain of impact velocities less than 100 mm/s is categorized as the quasi-static regime. The phenomenon of resistance provided by surfaces in such regime is termed as quasi-static robustness, and the corresponding surfaces are called quasi-statically robust. The current work aims at establishing the range of surface parameters required to ascertain quasi-static robustness.

### 1.2. Metastable Cassie state: geometric orientation

For a quasi-static deposition, the LA interface of the drop comes into contact with the apex of the surface roughness (Cassie state). The transition of a LA interface to the metastable Cassie state is governed by the mutual free energy values of the possible wetting states (Cassie, Wenzel and metastable Cassie). The transition from one wetting regime to the other occurs by one of two possible mechanisms, sag or depinning (figure 1). While the co-existence of both the mechanisms has been witnessed, we limit our discussion to cases where sag and depinning mechanisms are mutually exclusive. In the case of the sag mechanism, the apex of the roughness valleys pins the liquid, thereby causing a part of the LA interface to sag owing to gravitational force (figure 1a) [35, 36]. If the gravitational forces acting on the solid-liquid contact overcome the shear forces, the SL contact gets de-pinned from the apex of the roughness feature, and the corresponding mechanism is called depinning (figure 1b) [20, 37]. The metastable Cassie states attained via sag and depinning mechanisms are termed as sagged state and depinned state, respectively.

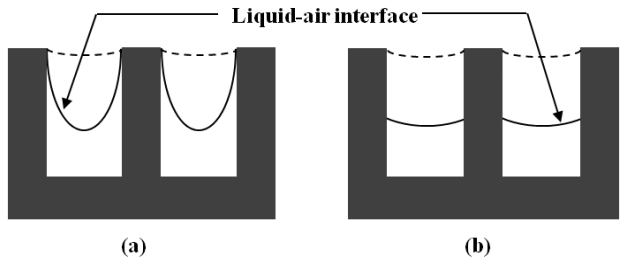

Fig.1 Mechanisms showing transition from Cassie to Wenzel state (a) Sag mechanism and (b) Depinning mechanism

## 2. Surface design for quasi-static robustness: Thermodynamic approach

Quasi-static robustness is given as the least likelihood of the LA interface to transition to a Wenzel state. Our discussion is limited to surface chemistries with Young's contact angle $\theta_Y > 90°$. The Cassie state, metastable Cassie states (multiple for various penetration depths and different configurations of the LA interface) and the Wenzel state can be sequentially encountered as the LA interface penetrates the roughness valley. At constant temperature and pressure, the droplet starts with the Cassie state and settles at the wetting state with the lowest free surface energy. It is imperative to determine the thermodynamic feasibility of a depinned state or a sagged state [36]. Additionally, the relative values of free surface energy need to be considered for each wetting state with respect to another wetting state. A case study is performed, wherein the mutual free energy values of the Cassie, the metastable Cassie and the Wenzel state are compared. If the free energies of the Cassie ($G_{CB}$), the metastable Cassie ($G_M$) and the Wenzel state ($G_W$) assume distinct values, six cases can be distinguished (table 1).

Table 1 Six combinations of free surface energy and their feasibility

| Case | Inequality correlations | Feasibility | Transition mode | Final state |
| --- | --- | --- | --- | --- |
| I | $G_{CB} < G_W < G_M$ | ✓ | No transition | Cassie |
| II | $G_{CB} < G_M < G_W$ | ✓ | No transition | Cassie |
| III | $G_M < G_{CB} < G_W$ | | | |
| IV | $G_W < G_{CB} < G_M$ | ✓ | Sag/Depinned | Wenzel |
| V | $G_W < G_M < G_{CB}$ | | | |
| VI | $G_M < G_W < G_{CB}$ | | | |

For a sagged state, pinning of the liquid-air interface under the drop results in a rise of liquid-air interfacial area and free surface energy. For a depinned state and $\theta_Y > 90°$, the liquid occupies the side-walls of the surface topology, thereby raising the total free surface energy. Hence, the surface energy of any metastable Cassie state (depinned or sagged) is higher than that of the Cassie state. Of these six cases, three cases (III, V and VI) advocate $G_{CB} > G_M$ (table 1), which is implausible. Hence, this discussion is limited to the remaining cases (I, II and IV), which involve $G_{CB} < G_M$. Cases I and II are characterized by a Cassie state that is energetically favorable in comparison to a Wenzel state, i.e. $G_{CB} < G_W$ (figure 2i and figure 2ii). On the other hand, case IV consists of an energetically favorable Wenzel state, i.e. $G_W < G_{CB}$ (figure 2iii). It has been proven that systems with an energetically favorable Wenzel state exhibit both sag transitions and depinning transitions [38].

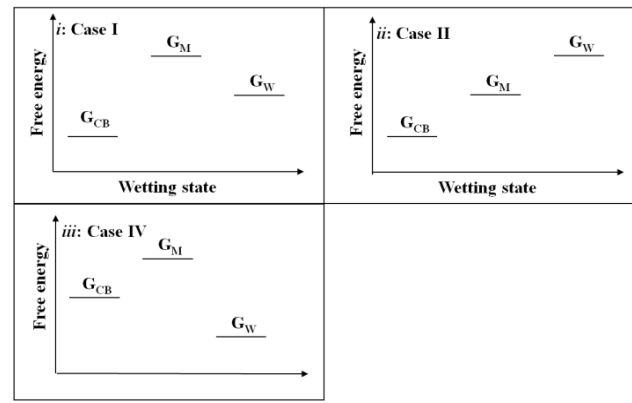

Fig.2 Role of the metastable Cassie state in determining robustness (i) Case I (ii) Case II (iii) Case IV

Here, efforts are made to investigate the conditions under which each of these cases renders quasi-static robustness. A square pillar surface topology of micrometer dimensions has been chosen as a template surface, with post width $a$ μm, post spacing $b$ μm and post height $c$ μm (figure 3i). In our previous work, a bottom-up approach had been employed, wherein the free surface energy of the system was determined in terms of the simplest building block, defined as a unit [11].

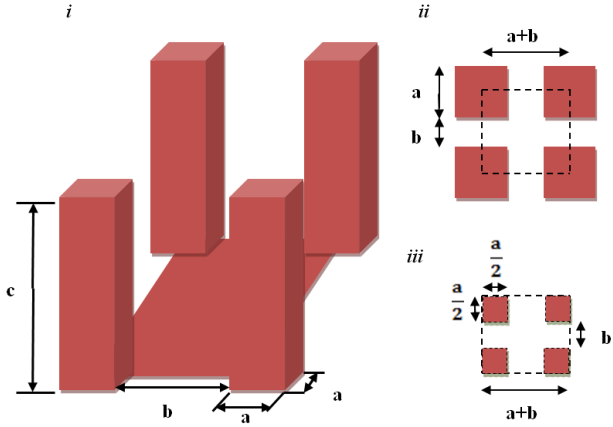

Fig.3: Square post geometry (i) 3 Dimensional view showing post width $a$ μm, post spacing $b$ μm and post height $c$ μm (ii) Top view of four nearest pillars, which outline a unit (iii) unit, or a simplest building block, i.e. unit, inscribed in the dotted rectangle, which is characterised by a post width $a/2$ μm, post spacing b μm and post height $c$ μm[11]

A wetting state $i$ can be denoted by a signature wetting parameter $j_i$, which is a unique footprint of the wetting state for a given surface topology. The apparent contact angle ($\theta_i$) is used to express the contact angle for the metastable depinned Cassie state ($\theta_{Mdep}$), the metastable sag state ($\theta_{Msag}$), the Cassie state ($\theta_{CB}$) and the Wenzel state ($\theta_W$). The free surface energy of a unit $G_i^{unit}$ is expressed as a function of the wetting parameter and the apparent contact angle (equation 1).

$$G_i^{unit} = \gamma_{LA}(a+b)^2 \left[\frac{2}{1+\cos\theta_i} + 1 + j_i\right] \quad (1)$$

For the Cassie and Wenzel states, the minimization of the available surface energy (for all the units under the drop) renders an explicit correlation between the apparent contact angle ($\theta_i$) and the wetting parameter ($j_i$) (equation 2). In the process of carrying out the energy minimization calculation for the sagged state, we show that the same correlation also applies to a sagged state (supporting information A).

$$\forall i \, \varepsilon \{CB, W, M_{sag}\}; \cos\theta_{\alpha i} = -1 - j_i \quad (2)$$

The wetting parameters and the expressions for APCA for each wetting state are listed in table 2.

Table 2 APCA ($\theta_i$) for various wetting states

| Wetting state | $\cos\theta_i$ |
|---|---|
| Cassie | $\cos\theta_{CB} = \left(\frac{a}{a+b}\right)^2 (1+\cos\theta_Y) - 1$ |
| Wenzel | $\cos\theta_W = \cos\theta_Y \left(1 + \frac{4ac}{(a+b)^2}\right)$ |
| Depinned | $\cos\theta_{Mdep} = func(a, b, \theta_Y, h)$ |
| Sagged | $\cos\theta_{Msag} = \left(\frac{a}{a+b}\right)^2 (1+\cos\theta_Y) + \frac{b^2 - b\sqrt{b^2+4c^2}}{(a+b)^2} - 1$ |

To analyse cases I, II and IV, the free surface energy values are calculated for distinct pairs of wetting states (denoted by subscripts $i$ and $j$). The difference in free surface energy is converted to a non-dimensional form by division with the product of the liquid-air interfacial tension and the area of the unit (equations 3 and 4).

$$\Delta G_{i,j}^{unit} = G_j^{unit} - G_i^{unit} \quad (3)$$

$$\Delta G_{i,j}^* = \frac{\Delta G_{i,j}^{unit}}{\gamma_{LA}(a+b)^2} \quad (4)$$

The sign of $\Delta G_{i,j}^*$ is crucial in the analysis of cases I, II and IV. For the depinned state, the free energy is a function of penetration depth. All the other wetting states (Cassie, Wenzel, sagged) share the same expression for the free surface energy. The free energy of a wetting state $i$ is a monotonically increasing function of the APCA ($\theta_i$) [12]. Thus, a comparative study of the free surface energy of two distinct wetting states $i$ and $j$ can be carried out by comparing the APCAs of the corresponding wetting states (equation 5).

$$\frac{|\Delta G_{i,j}^*|}{\Delta G_{i,j}^*} = \frac{|\theta_j - \theta_i|}{\theta_j - \theta_i}; \forall i,j \in (M_{sag}, CB, W) \ni i \neq j \quad (5)$$

For the depinned state, the APCA depends on the penetration depth. The analysis of a depinned state is carried out using free surface energy minimization and mechanics, as will be shown in the following section.

### 2.1. Robustness with an energetically favorable Cassie state (cases I and II)

Cases I and II can be classified by a common inequality ($G_{CB} < G_W$) (table 1). Since the Cassie state assumes a lower free surface energy, the Cassie contact angle is less than the Wenzel contact angle (equation 6).

$$\Delta G_{CB,W}^* \geq 0; \theta_{CB} \leq \theta_W \quad (6)$$

On substituting the cosines of $\theta_{CB}$ and $\theta_W$ (table 2), the maximum permissible value is obtained for the pillar spacing to width ratio, also termed as critical spacing to width ratio $(b/a)_{critical}$ (equation 7). Any b/a ratio exceeding this limit will lead to a transition to the Wenzel state.

$$\frac{b}{a} \leq \left(\frac{b}{a}\right)_{critical} = \sqrt{1 - \frac{4c\cos\theta_Y}{a(1+\cos\theta_Y)}} - 1 \quad (7)$$

Existing approaches at surface design involve b/a ratio less than the critical limit [12, 17, 19, 21, 39, 40] Although the critical limit is well-known, the domains of dependent parameters, namely c/a ratio and $\theta_Y$ have not yet been investigated. It can be shown that the critical limit assumes a positive real value, if and only if the Young's contact angle exceeds 90° (equation 8).

$$\forall \left(\frac{b}{a}\right)_{critical} > 0; \theta_Y \geq 90° \quad (8)$$

Alternatively, a minimum value of $\theta_Y$ ($\theta_{min}$) can be determined in order to have a feasible Cassie state (equation 9).

$$\theta_Y = \theta_{min} \geq \sec^{-1}\left(\frac{4\frac{c}{a}}{1-\left(1+\frac{b}{a}\right)^2} - 1\right) \quad (9)$$

$\theta_{min}$ monotonically increases with the b/a ratio and shows a monotonic fall with the c/a ratio (Figure 4). For surface conditions with b/a > 2 and c/a < 3, $\theta_{min}$ exceeds 120°. Since no polished surface has been reported with a $\theta_Y$ exceeding 120°, surface design beyond this range proves to be a daunting task.

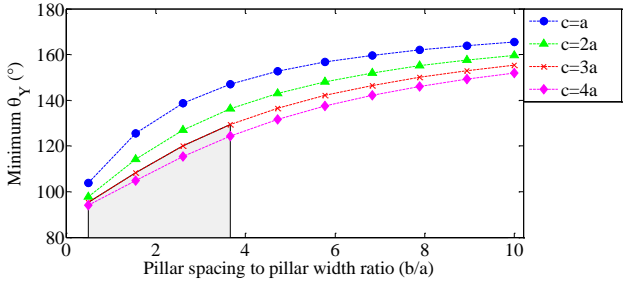

Fig.4: Minimum $\theta_Y$ vs. the ratio of pillar spacing to pillar width for different pillar heights

In recent times, $\theta_Y$ as high as 154.6° have been reported, in the process of incorporating nanostructures on micropillars [17, 41, 42]. Since the nanostructures are vanishingly small in comparison to micropillars, the APCA of such a modified surface has been approximated as its $\theta_Y$. While this process helps to achieve a suitable $\theta_{min}$, additional surface treatments are necessary, with consequent expenditures. Without such a surface treatment, the surface design will be limited to a very narrow range of surface parameters (b/a < 2, c/a > 3, as shown in the shaded area, Figure 4). Although a stable Cassie state is found with cases I and II, flexibility in the choice of surface parameters is accompanied by additional costs of surface design. In contrast, the analysis of case IV marks a brand new attempt at providing higher degree of user flexibility without expenditure.

### 2.2. Metastable state acts as an energy barrier (case IV)

In general for case IV b/a ratios exceed the critical limit, thus, the latter must assume a real positive value. In essence, case IV shares the necessary condition with cases I and II, i.e. a positive real value for the critical *b/a* ratio (equation 8). The metastable Cassie state acts as an energy barrier in the transition to a Wenzel state. Both depinning and sag transitions are possible [38]. For a surface with $\theta_Y > 90°$, the surface energy of a metastable Cassie state (depinned/sagged) is always greater than that of the Cassie state (equation 10). Thus, a surface is quasi-statically robust if the contact angles corresponding to depinning/sag transitions hold mathematically permissible values.

$$\Delta G_{C,M}^* > 0;\ \theta_M \in (0°, 180°) \tag{10}$$

#### 2.2.1. Sagged state

In the following, cases are identified where the APCA corresponding to the sagged metastable state hold a realizable value.

$$-1 \leq \cos\theta_{Msag} \leq 1 \tag{11}$$

Inequality 11 is explicitly expressed in terms of the b/a ratio (equation 12) (for derivation refer to supporting information A). It is seen that the b/a ratio is limited by a unique function of *c/a* ratio and $\theta_Y$. The upper limit is termed as sagged spacing to width ratio $(b/a)_{sag}$.

$$\frac{b}{a} \leq \left(\frac{b}{a}\right)_{sag} = \frac{(1+\cos\theta_Y)}{\sqrt{4\frac{c^2}{a^2} - 2(1+\cos\theta_Y)}} \tag{12}$$

For the fulfillment of case IV, the *b/a* ratio must be bounded by the critical and the sagged limits. Permissible values of the remaining parameters, namely c/a ratio and $\theta_Y$ are determined by elucidating the conditions, wherein the sagged *b/a* ratio exceeds the critical *b/a* ratio (equation 13).

$$\left(\frac{b}{a}\right)_{sag} - \left(\frac{b}{a}\right)_{critical} =$$

$$= \frac{(1+\cos\theta_Y)}{\sqrt{4\frac{c^2}{a^2} - 2(1+\cos\theta_Y)}} - \sqrt{1 - \frac{4c\cos\theta_Y}{a(1+\cos\theta_Y)}} + 1 > 0 \tag{13}$$

The difference between the sagged limit and the critical limit is plotted against *c/a* ratios for various values of $\theta_Y$ (Figure 5). We identify the magnitudes of *c/a* ratio and $\theta_Y$ that jointly result in a positive value for the difference. It is seen that the sagged limit exceeds the critical limit for 90° < $\theta_Y$ < 105°, and 0.75 < *c/a* < 0.9 (equation 14).

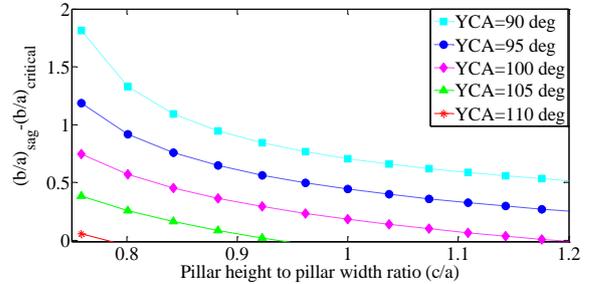

Fig.5 Identification of the domain of permissible parameters for case IV (sag)

$$0.75 < \frac{c}{a} < 0.9;\ 90° < \theta_Y < 105° \tag{14}$$

The range of $\theta_Y$, necessary for the fulfilment of case IV is significantly lower than that required for cases I and II. Hence, knowledge of the current analysis offers a higher degree of user flexibility with the choice of $\theta_Y$ without involving additional expenditures related to surface modification.

#### 2.2.2. Depinned state

For the fulfillment of case IV with a depinning transition, the APCA corresponding to a depinned state ($\theta_{Mdep}$) should assume mathematically realizable values (equation 15).

$$-1 \leq \cos\theta_{Mdep} \leq 1 \tag{15}$$

Thus, to pinpoint the surface parameters, the cosine of $\theta_{Mdep}$ needs to be explicitly expressed in terms of the surface parameters (*a, b, c, $\theta_Y$*). The penetration depth (*h*) shares an implicit correlation with the $\theta_{Mdep}$, and is given as the characteristic set of equations [11]. Several algebraic expressions that constitute the equation contain fractional exponents of $\theta_{Mdep}$. Using binomial expansion for fractional exponents, these expressions are converted to linear functions of $\theta_{Mdep}$ (supporting information B). Upon expansion, the characteristic set of equations is expressed as a quadratic equation of $\theta_{Mdep}$. For a mathematically realizable $\theta_{Mdep}$, the quadratic equation must have a positive discriminant (necessary condition) and at least one real root with a value between -1 and 1 (equation 15, sufficient condition). It is seen that the necessary condition and the sufficient condition are mutually exclusive for any set of surface parameters. Hence, it is safe to infer that the minimization of free surface energy is not sufficient to analyze the thermodynamics of a depinned state for $\theta_Y > 90°$. Since a wetting state is a direct consequence of how the liquid comes into contact with the surface, the origin of a depinned state is traced back to the kinematics of a transitioning LA interface (discussed in chapter 3).

Thus, the problems with surface design pertaining to cases I and II are discussed, and the flexibility of surface design is introduced for case IV. The accurate ranges provided for $c/a$ and $\theta_Y$ should invite the attention of surface designers. The inability of surface energy minimization alone to explain a depinning transition forms the prelude to understanding robustness as a dynamic problem.

Table 3 Necessary and sufficient conditions for quasi-static robustness, and the allowed values of surface chemistry ($\theta_Y$)

| Case | Guidelines for Surface design |
| --- | --- |
| I, II | $\forall \theta_Y > 90°$ <br> $\frac{b}{a} \leq \left(\frac{b}{a}\right)_{critical}$ |
| IV (sag only) | $90° < \theta_Y < 105°$ <br> $0.75 < \frac{c}{a} < 0.9$; <br> $\left(\frac{b}{a}\right)_{critical} \leq \frac{b}{a} \leq \left(\frac{b}{a}\right)_{sag}$ |

## 3. Surface design for quasi-static robustness: Pressure balance approach

As previously highlighted, we argued that solely the minimization of surface energy is insufficient to quantify a depinned state for $\theta_Y > 90°$. We visualize the depinned state in terms of the drop-surface kinematics, i.e. the forces experienced by the drop and the surface. First, it is imperative to prove that the depinned state occurs as a result of quasi-static deposition. As stated before, it is very difficult for a user to distinguish a quasi-static drop deposition (corresponding to impact velocities less than 100 mm/s) from drop deposition with virtually no impact velocity. Thus, an impact velocity less than 100 mm/s can be present in a static contact angle measurement (CAM), which can go unnoticed by the user. Therefore, static CAM can be categorized as quasi-static deposition [32-34]. Thus, if a depinned state is proved to be existent with static CAM, the depinned state must also exist with quasi-static deposition. The existence of a depinned state with static CAM can be determined by noticing the departure of an experimentally recorded APCA from the Cassie contact angle. In the process of investigating reported CAMs in literature, for surfaces with $\theta_Y > 90°$, Erbil et al. conducted a detailed mathematical analysis of the deviation of APCA from that predicted by Wenzel and Cassie equations [5, 43, 44]. APCAs for 28 different surfaces (with known $\theta_Y$ and square pillar geometry) were listed and, using the Cassie equation, converted to a solid fraction term. If a solid fraction, calculated from the measured APCA with known $\theta_Y$ and square pillar geometry, exceeded the geometric solid fraction corresponding to the Cassie state, penetration and consequently, a depinned state could be inferred. It was found that 10.7 % of the surfaces correspond to a finite penetration depth, and thus, a depinned state. Given the magnitude of the above percentage, the results are too insignificant to confirm a depinned state. However, in these calculations it is assumed that the liquid-air fraction and the solid fraction add up to unity, which is incorrect for a depinned state [6].

Under the assumption that the liquid-air fraction is independent of the penetration depth and the solid fraction, we repeat the mathematical steps of Erbil (supporting information C). It is seen that the solid fraction, determined from APCA exceeds the geometric solid fraction in 89 % of the surfaces. This observation acts as evidence for the occurrence of finite penetration with static CAM, and consequently the existence of a depinned state. Since static CAM is categorized as a quasi-static deposition, it is safe to infer that a depinned state is a possible outcome of a quasi-static deposition. Current literature lacks a kinematic approach toward the quantification of a depinned state. It is a daunting task to express the APCA for a depinned state ($\theta_{Mdep}$) in terms of the force experienced or the pressure acting on the surface. The APCA is determined by minimization of the total available free energy, i.e. the work done by pressure terms acting on the system and the surface energy. While surface energy can be calculated with knowledge of surface geometry and chemistry, the net pressure imparted by a drop on a surface is contingent to the experimental conditions, namely pressure of compressed air underneath the droplet, the pressure exerted by the drop on the surface and relative humidity. Thus, characterisation of a depinned state, i.e. determination of penetration depth requires the consideration of pressure acting on the surface. While it is difficult to establish a direct correlation between $\theta_{Mdep}$ and the drop velocity ($v$), we seek to express the penetration depth ($h$) in terms of $v$. A pressure balance is carried out, which lists the pressure acting on the surface, i.e. the wetting pressure ($P_{wetting}$) and the pressure exerted by the surface, i.e. the antiwetting pressure ($P_{antiwetting}$).

The antiwetting pressure ($P_{antiwetting}$) corresponds to the energy difference between the homogeneous and the heterogeneous wetting regimes [16, 17, 40]. The antiwetting pressure denotes the force per unit area offered to a water droplet as it transitions from a depinned metastable Cassie state to a Wenzel state. First, the force acting on the LA interface ($F_{antiwetting}$) is calculated by measuring the rate of change of the free energy $\Delta G_{CB,M}{}^{unit}$ with respect to penetration depth ($dh$) (equation 16).

$$F_{antiwetting} = \frac{d(\Delta G_{CB,M}{}^{unit})}{dh} = -4\gamma_{LA} a \cos\theta_Y \quad (16)$$

Next, the corresponding pressure term is calculated by division with the area of the LA interface in a unit ($A_{LA}$) (equation 17).

$$P_{antiwetting} = \frac{1}{A_{LA}} \frac{d(\Delta G_{CB,M}{}^{unit})}{dh} = -\frac{4\gamma_{LA} a \cos\theta_Y}{b(2a+b)} \quad (17)$$

On the other hand, there exist two independent and unique definitions for the wetting pressure ($P_{wetting}$). The first definition of wetting pressure attributes wetting pressure to the drop weight and drop curvature, and is associated with a static drop dispense ($P_{wetting,static}$) [18, 40, 45]. This static wetting pressure constitutes the

Laplace pressure and hydrostatic pressure (equation 18). For drops with radii less than the capillary length, the hydrostatic pressure is negligible.

$$P_{Wetting,static} = P_{Laplace} + P_{hydrostatic} \cong P_{Laplace} \quad (18)$$

$$P_{Laplace} = \frac{2\gamma_{LA}}{R} \quad (19)$$

The second definition of wetting pressure involves the drop kinetic energy and the shock-wave formed as a result of drop-surface impact, and corresponds to velocity driven wetting ($P_{wetting,dynamic}$) [18, 46]. The dynamic wetting pressure is a sum of Bernoulli pressure ($P_{Bernoulli}$) and Water hammer pressure ($P_{WH}$) (equation 20).

$$P_{wetting,dynamic} = P_{WH} + P_{Bernoulli} \quad (20)$$

Bernoulli pressure ($P_{Bernoulli}$) denotes the ratio of the kinetic energy of the impacting droplet to its volume (equation 21).

$$P_{Bernoulli} = 0.5\,\rho v^2 \quad (21)$$

The water hammer pressure ($P_{WH}$) corresponds to the shock-wave precisely at the moment of impact and is known to be sufficiently strong to cause a wetting transition at low velocities [47] (equation 22).

$$P_{WH} = k\rho c_1 v \quad (22)$$

Here, $c_1$ denotes the speed of sound in water (1497 ms$^{-1}$). The coefficient k refers to a collision factor which describes the elasticity of the collision. The value of k approaches a maximum value of 0.5 for nearly elastic collisions [48]. The experiments on droplet impingement typically use a droplet speed of the order of ms$^{-1}$, for which the water hammer coefficient is typically approximated as 0.2 [20, 22, 49]. For low velocities and high droplet volumes, the collision is known to be inelastic, which lowers the coefficient k to the order of 0.001 (equation 23) [18]. Experimentally determined values of the coefficient, as available in literature, range between 0.1 and 0.001 [18, 46].

$$k = f(v, V);\ \frac{dk}{dv} > 0;\ \frac{dk}{dV} < 0 \quad (23)$$

Recently, Dash et al. have developed an empirical relation between the water-hammer coefficient and the anti-wetting pressure, which has been expressed with no loss of generality for surfaces with grooves as well as pillars [42, 46].

$$k = 2.57 \frac{P_{antiwetting}}{Nm^{-2}} 10^{-7} + 7.53\,10^{-4} \quad (24)$$

Substituting the value of the water-hammer coefficient ($k$), the dynamic wetting pressure can be explicitly expressed in terms of the impact velocity (equation 25).

$$P_{wetting,dynamic} =$$
$$= (2.57 \frac{P_{antiwetting}}{Nm^{-2}} 10^{-7} + 7.53\,10^{-4})\rho c_1 v + 0.5\,\rho v^2 \quad (25)$$

Now, if the wetting pressure exceeds the antiwetting pressure, the LA interface will penetrate the apex of the surface roughness. A parameter is coined, namely wetting state determining depth ($h_{WSDD}$), which correlates the wetting state to the velocity. The aforementioned factor can be understood as the height of a column of water, which generates a hydrostatic pressure, which is identical to the numerical difference between the wetting and the antiwetting pressures (equation 26). The parameter can be expressed in terms of the surface chemistry ($\theta_Y$), surface geometry ($a,b,c$) and the velocity of impact ($v$).

$$\forall l \in \{static, dynamic\};\ h_{WSDD,l} = \frac{P_{wetting,l} - P_{antiwetting}}{\rho g} \quad (26)$$

The suffix $l$ refers to the mode of drop deposition, i.e. static/dynamic. Any $h_{WSDD,l}$ which exceeds the pillar height corresponds to the Wenzel state, while any negative $h_{WSDD,l}$ implies a Cassie state (table 4). Only values of $h_{WSDD,l}$ which fall in between zero and the pillar height $c$ imply a depinned metastable Cassie state.

Table 4 Wetting state determining depth ($h_{WSDD,l}$)

| $h_{WSDD,l}$ (μm) | Penetration | Wetting state |
| --- | --- | --- |
| $h_{WSDD,l} \leq 0$ | No penetration | Cassie |
| $0 < h_{WSDD,l} \leq c$ | Partial penetration | Depinned |
| $h_{WSDD,l} > c$ | Complete penetration | Wenzel |

For a surface with known geometric parameters ($a$, $b$, $c$), chemistry ($\theta_Y$), and a drop with known chemistry (LA surface tension) and drop volume (expressed in terms of radius), the static wetting state determining depth ($h_{WSDD,static}$) is calculated using equations 18, 19 and 26 (equation 27).

$$h_{WSDD,static} = \frac{2\gamma_{LA}}{R} + \frac{4\gamma_{LA}a\cos\theta_Y}{\rho g b(2a+b)} \quad (27)$$

The dynamic wetting state determining depth ($h_{WSDD,dynamic}$) is found by substituting equations 17 and 25 into equation 26 (equation 28). It depends on the surface parameters ($a$, $b$, $c$, $\theta_Y$), the liquid properties (density, LA surface tension) and the velocity of impact.

$$h_{WSDD,dynamic} =$$
$$= 7.53\,10^{-4} \frac{c_1 v}{g} + 0.5\,\frac{v^2}{g} + \frac{4\gamma_{LA}a\cos\theta_Y}{\rho g b(2a+b)}\left(1 + \frac{2.57\,10^{-7}}{Nm^{-2}} \cdot \frac{cv}{g}\right) \quad (28)$$

It is assumed that the deposited liquid drop retains its spherical shape. The height of a pillar should be such that partial or complete penetration of water should not give rise to any change in the configuration of the deposited drop. Pillar heights ($c$) lower than 300 μm ensure that the volume of the droplet under the roughness features does not contribute significantly to the total drop volume [11]. Thus, a pillar height no greater than 300 μm is chosen. For $\theta_Y$ of 100° and a pillar width of 15 μm, the dependence of the wetting state determining depth ($h_{WSDD,dynamic}$) of water (surface tension: 0.072 N/m, density: 1000 kg/m$^3$) with increasing pillar spacing is investigated exemplarily for four different velocities, namely 20 mm/s, 50 mm/s, 75 mm/s and 100 mm/s (figure 6).

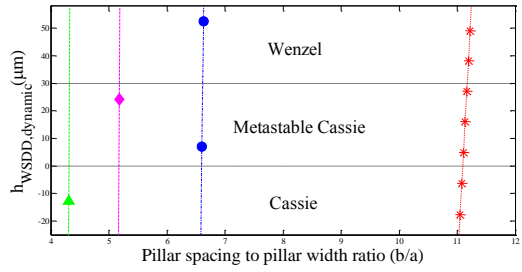

Fig.6: Wetting state determining depth ($h_{WSDD,dynamic}$) representing the cases of Cassie, depinned and Wenzel state for $\theta_Y = 100°$, $a = 15$ μm and $c = 30$ μm. Four sets of velocities are used: 20 mm/s (red asterisk), 50 mm/s (blue circles), 75 mm/s (pink diamonds), 100 mm/s (green triangles)

The antiwetting pressure exhibits an inverse square relationship with the spacing to width ratio, and hence falls sharply with increasing spacing to width ratios. For an impact velocity of 20 mm/s, $a = 15$ μm, $c = 30$ μm, Cassie and Wenzel states are encountered at $b/a = 11$ and $b/a = 11.3$ respectively (figure 6). This means that only for $11 < b/a < 11.3$, the liquid partially impales into the roughness valleys, thereby generating a depinned metastable Cassie state.

### 3.1. Surface chemistry ($\theta_Y$)

The evolution of wetting state determining depth with surface geometry is further investigated by considering two different surface chemistries. Two values of $\theta_Y$ are chosen as a hypothetical template, namely 100° and 120°. For each of the four above mentioned velocities, i.e. 20 mm/s, 50 mm/s, 75 mm/s and 100 mm/s, the wetting state determining depth is plotted (figure 7). The plots denote the transition of the LA interface past the apex of the roughness features. With a rise in $\theta_Y$, the wetting transition from Cassie starts at a higher value of pillar spacing. For a velocity of 75 mm/s, the transitions with $\theta_Y = 100°$ and $\theta_Y = 120°$ occur at $b/a = 5.2$ and 9.4, respectively.

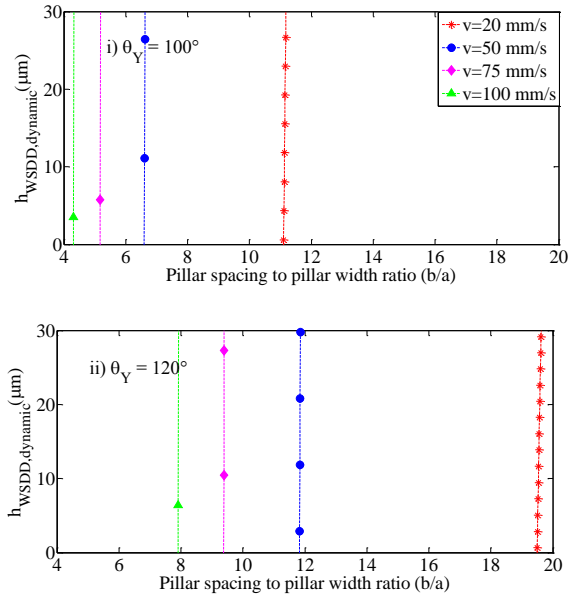

Fig.7 Variation of the wetting state determining depth for water for surface with $\theta_Y=100°$ (i) and $\theta_Y=120°$ (ii)

### 3.2. Pillar spacing to width ratio-quasi-static limit

As the LA interface starts to transition from a Cassie state, the wetting state determining depth ceases to be negative. We investigate the applicability of the dynamic mode of drop deposition in the process of revisiting the static CAM for 4 sets of experimental data available in literature [3, 4, 40]. The experiments are chosen in manner such that the CAM results correspond to surfaces with both square pillar and cylindrical micropillars, hence avoiding any loss of generality. Each set of surfaces is marked by a given chemistry, a fixed pillar width and varying pillar spacing to width ratios (table 5).

The experimentally reported spacing to width ratio corresponding to a wetting transition ($(b/a)_{exp}$) is recorded for the above sets of experiments. An accurate depiction of the dynamic model (equation 28) would yield no penetration ($h_{WSDD}=0$) for a spacing to width ratio equalling the experimentally reported value (equation 29). Thus, at the onset of a wetting transition, the dynamic wetting pressure equals the antiwetting pressure, and is given as the calculated wetting pressure ($P_{wetting,dynamic}^{calc}$).

$$\forall \frac{b}{a} = \left(\frac{b}{a}\right)_{exp} ; P_{wetting,dynamic}^{calc} = P_{antiwetting} \qquad (29)$$

The dynamic wetting pressure ($P_{wetting,\ dynamic}$) can be expressed in terms of the antiwetting pressure (equation 25). Using the expression for the water hammer coefficient, equation 29 is simplified to render impact velocities ($v_{calc}$) for antiwetting pressures (equation 30) corresponding to the square (equation 31) and cylindrical micropillars (equation 32).

$$v_{calc} = \frac{g}{c_1} \frac{P_{antiwetting}}{\left(7.53\ 10^{-4} - P_{antiwetting}\left(\frac{2.57\ 10^{-7}}{Nm^{-2}}\right)\right)} \qquad (30)$$

Square micropillars:

$$v_{calc} = \frac{g}{c_1} \frac{4\ \gamma_{LA}|\cos\theta_Y|}{a((1+(\frac{b}{a})_{exp})^2-1)\left(7.53\ 10^{-4} - \frac{4\ \gamma_{LA}|\cos\theta_Y|}{a((1+(\frac{b}{a})_{exp})^2-1)}\left(\frac{2.57\ 10^{-7}}{Nm^{-2}}\right)\right)} \qquad (31)$$

Cylindrical micropillars :

$$v_{calc} = \frac{g}{c_1} \frac{\pi\ \gamma_{LA}|\cos\theta_Y|}{a((1+(\frac{b}{a})_{exp})^2-1)\left(7.53\ 10^{-4} - \frac{\pi\ \gamma_{LA}|\cos\theta_Y|}{a((1+(\frac{b}{a})_{exp})^2-1)}\left(\frac{2.57\ 10^{-7}}{Nm^{-2}}\right)\right)} \qquad (32)$$

Table 5 clearly shows that calculated impact velocities do not exceed 100 mm/s, i.e. they fall well within the quasi-static regime. The above finding validates the applicability of the dynamic pressure consideration in understanding the quasi-static drop dispense, and consequently, the depinned transition.

Table 5 Calculated impact velocity corresponding to wetting transition for 4 sets of surfaces

| Surface | $\theta_Y$ (°) | $a$ (μm) | $(b/a)_{exp}$ | $(b/a)_{critical}$ | $v_{calc}$ (mm/s) |
|---|---|---|---|---|---|
| Varanasi et al. [40] | 120 | 15 | 7.6 | 1.77 | 100 |
| Barbieri et al. [3] | 110 | 10 | 11 | 2.05 | 46 |
| Bhushan et al.[4] | 109 | 5 | 12 | 1.205 | 87 |
| | 109 | 14 | 10.85 | 1.27 | 32 |

For all sets of surfaces, $(b/a)_{exp}$ is significantly higher than $(b/a)_{critical}$ (table 5). Thus, for impact velocities corresponding to the quasi-static regime, it is possible to evade a Wenzel state even with $b/a$ ratios well exceeding the critical limit. We propose that a quasi-statically robust surface should possess an antiwetting pressure such that any velocity in the quasi-static regime can be withstood. Since the minimum known impact velocity is about 100 mm/s, the maximum velocity that can be unknowingly imparted during dispense is chosen to be 100 mm/s. Hence, the antiwetting pressure must exceed the dynamic wetting pressure corresponding to an impact velocity of 100 mm/s. Since the Bernoulli pressure is insignificant in this velocity regime, equation 25 is modified by excluding the same (equation 33).

$$P_{antiwetting} \geq \rho c_1 v (2.57 \frac{P_{antiwetting}}{Nm^{-2}} \times 10^{-7} + 7.53 \times 10^{-4}) \tag{33}$$

Substituting the density of water (1000 kg/m$^3$), the speed of sound (1497 m/s) and the impact velocity at 100 mm/s, the minimum antiwetting pressure is found to be 117.23 Nm$^{-2}$ (equations 34-35).

$$P_{antiwetting} \geq \frac{7.53 \times 10^{-4}}{(\frac{1}{1000 \times 1497 \times 0.1 Nm^{-2}} - \frac{2.57}{Nm^{-2}} \times 10^{-7})} \tag{34}$$

$$P_{antiwetting} \geq 117.23 Nm^{-2} \tag{35}$$

Since the minimum limit for the antiwetting pressure has been simply deduced without involving the exact surface parameters in question, it applies to any surface topology without a loss of generality. For a square pillar surface, the minimum antiwetting pressure is substituted in equation 17 to provide an explicit correlation among surface parameters (supporting information D). In this process, it is seen that the $b/a$ ratio can possess a value not exceeding an upper boundary. We coin the upper boundary as the quasi-static spacing to width ratio $(b/a)_{QS}$ (equation 36).

$$(\frac{b}{a})_{critical} \leq \frac{b}{a} \leq (\frac{b}{a})_{QS} = \sqrt{1 - \frac{2456.64 \cos \theta_Y}{a}} - 1 \tag{36}$$

It is extremely important to pinpoint the domains of $a$, $c$ and $\theta_Y$ to fulfill equation 36. Thus, the current criterion is satisfied, when both the critical and quasi-static limits assume positive values, and the quasi-static limit exceeds the critical limit. Substitution of the individual magnitudes of the critical and quasi-static limits, followed by simplification renders the domain of $a$, $c$ and $\theta_Y$. It is seen that $\theta_Y$ and the pillar height $c$ share a correlation (equation 37, supporting information D). The Young's contact angle $\theta_Y$ can assume any value not exceeding an upper boundary, as determined by pillar height. Since the cosine function is monotonically decreasing, higher pillar height is associated with a narrower set of options for $\theta_Y$.

$$90° \leq \theta_Y \leq \cos^{-1}(\frac{c}{614.16} - 1) \tag{37}$$

From the standpoint of surface energy minimization, a depinned metastable state is not plausible for a surface with $\theta_Y > 90°$. However, upon modifying an existing investigation of static CAM, it is found that 85 % of surfaces with $\theta_Y > 90°$ exhibit a penetration, and hence, a depinned state. Since impact velocities less than 100 mm/s (quasi-static deposition) have not been recorded in literature, it has been postulated that such velocities can be accidentally encountered during static CAMs. Using a kinetic approach, the wetting pressure and the antiwetting pressure acting on a solid surface are balanced. While the magnitude of wetting pressures correspond to experimentally verified expressions available in literature, the antiwetting pressures are obtained from a series of existing CAM experiments. Since the drop velocity corresponding to a wetting transition does not exceed 100 mm/s, it is proved that quasi-static deposition can enforce a depinned state, and thus, a wetting transition. Thus, a robust surface must withstand an impact velocity of 100 mm/s, which corresponds to an antiwetting pressure of 117.23 Nm$^{-2}$. For a square pillar surface, the surface characteristics are determined that lead to such an antiwetting pressure. It is found that the spacing to width ratio can assume values higher than the critical limit. Also, the corresponding $\theta_Y$ cannot exceed a maximum value determined by the pillar height. Hence, we provide quantitative evidence that it is possible to achieve robustness without very high values of $\theta_Y$ or a narrow range of $b/a$ ratios.

## 4. Conclusion

In general, design of a robust superhydrophobic surface comprises high Young's contact angle (typically exceeding 120°) and spacing to width ratios limited by a critical upper bound (typically less than 2). For several applications, robustness is sufficient for velocities no greater than 100 mm/s (quasi-static regime). We show that quasi-static robustness can be achieved with low values of Young's contact angle (less than 105°), and with b/a ratio exceeding the critical limit. Based on surface energy minimization, a case is pinpointed wherein, despite an energetically favorable Wenzel state, the sagged state acts as an energy barrier between the Cassie and Wenzel states. For a square pillar surface, such a case is found for a specific range of surface chemistry (90° < $\theta_Y$ < 105°) and pillar height to width ratios (0.75 < c/a < 0.9). For the above mentioned domain, robustness is possible with spacing to width ratios significantly higher than the critical limit. Additionally, robustness is investigated from the standpoint of mechanics, wherein the pressures acting on the drop-surface system are analyzed corresponding to the quasi-static deposition regime. From existing literature, static contact angle measurements on four sets of surfaces have been put forward to prove that wetting transitions are governed by dynamic pressure even at a quasi-static regime. For the first time, it is postulated that a surface, regardless of its topology or geometry, should offer a minimum antiwetting pressure of 117 Pa to be quasi-statically robust. In order to have such a pressure with a square pillar surface, the interdependence of pillar height and surface chemistry is clearly depicted. Thus, both from the standpoints of thermodynamics and classical mechanics, we prove that a wider choice of surface characteristics is available for quasi-static robustness than that existent in contemporary methods.

# Supplementary Information

**Title:** Design of a robust superhydrophobic surface: thermodynamic and kinetic analysis.

**Authors:** Anjishnu Sarkar, Anne-Marie Kietzig

## Supporting information A

### A1. Determination of the free energy and APCA for a sagged state

The free surface energy for a sagged state ($G_{Msag}^{unit}$) is expressed in terms of the surface chemistry ($\theta_Y$), interfacial tension ($\gamma_{LA}$), the LA interfacial area ($A_{LA}^{unit}$) and the SL interfacial area ($A_{SL}^{unit}$) (equations A1-A3). It is assumed that the LA interface is pinned to the center of the unit, thereby forming a square pyramid. The free surface energy can be reduced to a dimensionless form ($G_{Msag}^{*}$) (equation A4). The dimensionless free surface energy is expressed in terms of the APCA for the sagged state ($\theta_{Msag}$) and the wetting parameter ($j_{Msag}$){Sarkar, 2013 #1314}.

$$G_{Msag}^{unit} = \gamma_{LA}(A_{LA}^{unit} - A_{SL}^{unit} \cos\theta_Y) \qquad \text{A1)}$$

$$A_{LA}^{unit} = \frac{2(a+b)^2}{1+\cos\theta_{Msag}} + 2ab + b\sqrt{b^2 + 4c^2} \qquad \text{A2)}$$

$$A_{SL}^{unit} = a^2 \qquad \text{A3)}$$

$$G_{Msag}^{*} = \frac{G_{Msag}^{unit}}{\gamma_{LA}(a+b)^2} = \frac{2}{1+\cos\theta_{Msag}} + 1 + j_{Msag} \qquad \text{A4)}$$

Where

$$j_{Msag} = -\left(\frac{a}{a+b}\right)^2 (1+\cos\theta_Y) + \frac{b\sqrt{b^2+4c^2}-b^2}{(a+b)^2} \qquad \text{A5)}$$

Upon minimization of surface energy minimization, the wetting parameter $j_{Msag}$ can be directly correlated with $\theta_{Msag}$ (equation A6).

$$\frac{dj_{Msag}}{d\theta_{Msag}} = 0; \quad 1 + \cos\theta_{Msag} + j_{Msag} = 0 \qquad \text{A6)}$$

Using equations A5 and A6, an empirical relationship can be found for $\theta_{Msag}$ (equation A7).

$$\cos\theta_{Msag} = \left(\frac{a}{a+b}\right)^2 (1+\cos\theta_Y) - 1 + \frac{b^2+b\sqrt{b^2+4c^2}}{(a+b)^2} \qquad \text{A7)}$$

## A2. Domain of surface parameters for a thermodynamically feasible sagged state

For a sagged state to be thermodynamically feasible, $\theta_{Msag}$ should assume geometrically realizable values (equation A8). Equation A8 comprises two inequalities and is consequently simplified (equations A9-A12).

$$-1 \leq \cos\theta_{Msag} \leq 1 \tag{A8}$$

$$-1 \leq \left(\frac{a}{a+b}\right)^2 (1 + \cos\theta_Y) - 1 + \frac{b^2 - b\sqrt{b^2 + 4c^2}}{(a+b)^2} \leq 1 \tag{A9}$$

$$0 \leq \frac{a^2(1+\cos\theta_Y) + b^2 - b\sqrt{b^2 + 4c^2}}{(a+b)^2} \leq 2 \tag{A10}$$

$$0 \leq a^2(1 + \cos\theta_Y) + b^2 - b\sqrt{b^2 + 4c^2} \leq 2(a+b)^2 \tag{A11}$$

$$b\sqrt{b^2 + 4c^2} \leq a^2(1 + \cos\theta_Y) + b^2 \leq 2(a+b)^2 + b\sqrt{b^2 + 4c^2} \tag{A12}$$

Equation A12 is split into two inequalities (equations A13 and A14). Since the cosine of a function must be bounded by -1 and 1, both equations A13 and A14 must be correct.

$$a^2(1 + \cos\theta_Y) + b^2 \leq 2(a+b)^2 + b\sqrt{b^2 + 4c^2} \tag{A13}$$

$$b\sqrt{b^2 + 4c^2} \leq a^2(1 + \cos\theta_Y) + b^2 \tag{A14}$$

Equation A13 can be expressed as the sum of equations A15 and A16, which are individually true without any loss of generality. Hence, equation A13 is always correct.

$$a^2(1 + \cos\theta_Y) \leq 2a^2 \leq 2(a+b)^2 \tag{A15}$$

$$b^2 \leq b\sqrt{b^2 + 4c^2} \tag{A16}$$

Thus, the sagged state is feasible if and only if equation A14 is true. Equation A14 is squared and simplified to give the range of permissible spacing to width ratios (equations A17-A20). The spacing to width ratio is limited by a maximum value, here termed as sagged spacing to width ratio.

$$b^4 + 4b^2c^2 \leq b^4 + a^4(1 + \cos\theta_Y)^2 + 2b^2a^2(1 + \cos\theta_Y) \tag{A17}$$

$$4b^2c^2 \leq a^4(1+\cos\theta_Y)^2 + 2b^2a^2(1+\cos\theta_Y) \qquad \text{A18)}$$

$$b^2(4c^2 - 2a^2(1+\cos\theta_Y)) \leq a^4(1+\cos\theta_Y)^2 \qquad \text{A19)}$$

$$\frac{b}{a} \leq \left(\frac{b}{a}\right)_{sag} = \frac{(1+\cos\theta_Y)}{\sqrt{4\frac{c^2}{a^2} - 2(1+\cos\theta_Y)}} \qquad \text{A20)}$$

However, the sagged limit must exceed the critical limit for $\theta_Y > 90°$. The difference between the sagged limit and the critical limit is plotted with respect to the height to width ratio (c/a) for multiple surface chemistries (figure A1). For $\theta_Y > 105°$, the critical limit exceeds the sagged limit, and hence a feasible sagged limit cannot exist for the corresponding surface chemistries.

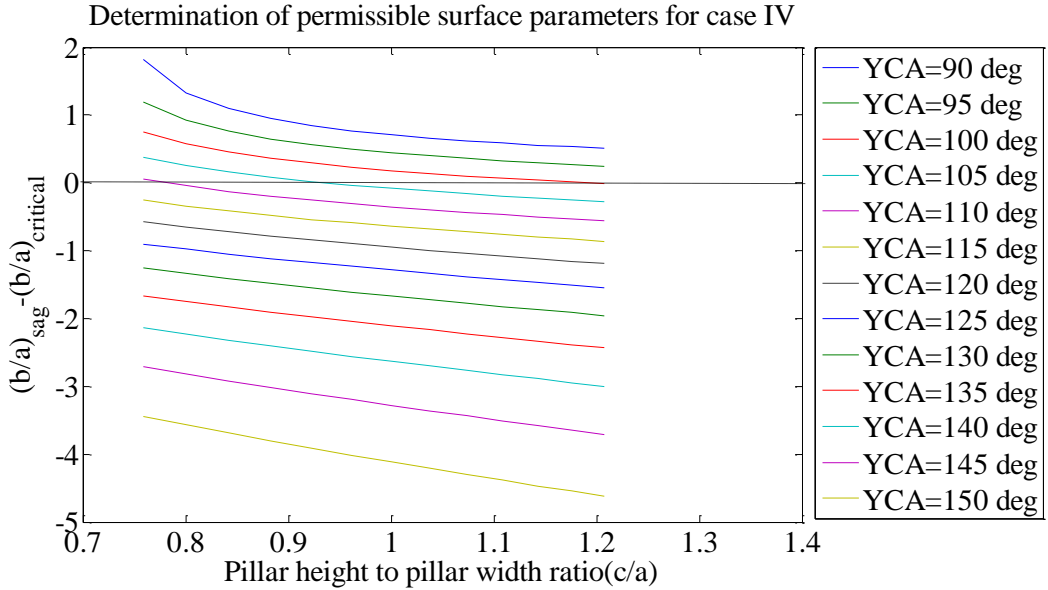

Figure A1 Variation of difference in sagged and critical limits with pillar height to width ratios

It is seen that the sagged limit assumes a real, positive value for a pillar height to width ratio greater than 0.7. The sagged limit exceeds the critical limit for $90° < \theta_Y < 105°$ (figure A2).

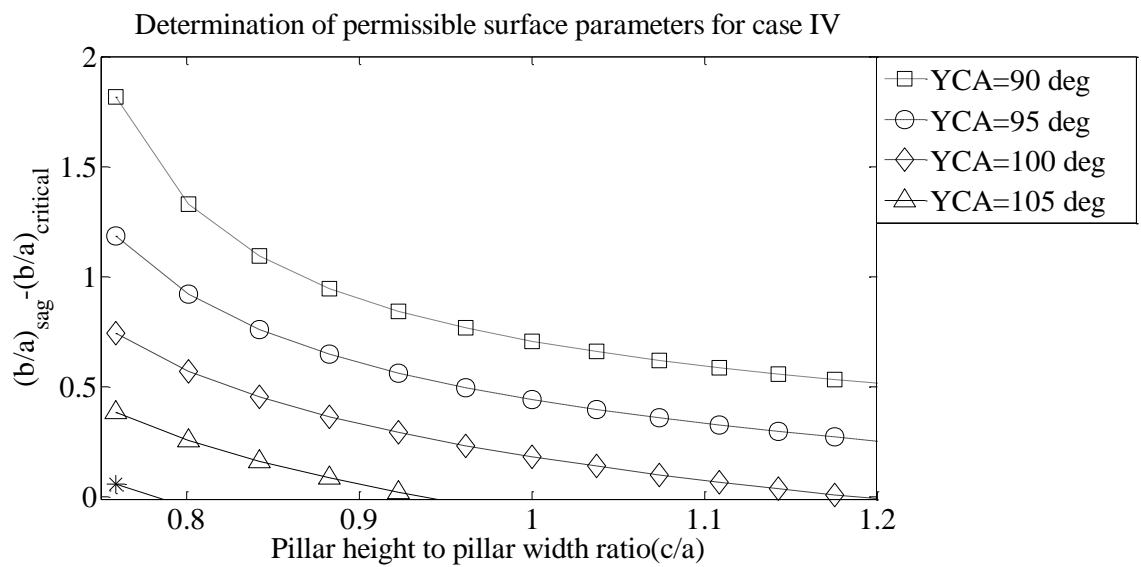

Figure A2 Domain of permissible height to width ratios and surface chemistry

**Supporting information B: General equation of wettability for square pillar geometry**

For a depinned state to exist, the APCA ($\theta_{Mdep}$) should hold appropriate values for surfaces with $\theta_Y > 90°$ (equation B1). The APCA of a depinned state shares an implicit correlation with the penetration depth, and is given as the characteristic set of equations (Sarkar and Kietzig 2013) (equation B2).

$$\forall \theta_Y \geq 90°; \; 0° \leq \theta_{Mdep} \leq 180°; \; -1 \leq \cos\theta_{Mdep} \leq 1 \qquad \text{B1)}$$

$$\frac{4ah\cos\theta_Y(1+\cos\theta_{Mdep})}{(a+b)^2} + \cos\theta_{CB}(1+\cos\theta_{Mdep}) - 2 + \varphi = 0 \qquad \text{B2)}$$

where $= (2+\cos\theta_{CB})^{\frac{1}{3}}(1-\cos\theta_{CB})^{\frac{2}{3}}(1-\cos\theta_{Mdep})^{\frac{1}{3}}(2+\cos\theta_{Mdep})^{\frac{2}{3}}$.

The aforementioned symbols are presented in Table 1. The expression $\varphi$ is a non-linear function of $\theta_{Mdep}$ and $\theta_{CB}$.

Table B1: Glossary of the symbols used

| Symbol | Description |
|---|---|
| $\theta_Y$ | Young's contact angle (YCA) |
| $a$ | Width of micrometer sized pillar |
| $b$ | Spacing between consecutive pillars |
| $h$ | Penetration depth of liquid in roughness valleys. |
| $\theta_{Mdep}$ | Apparent contact angle (APCA) corresponding to $h>0$ |
| $\theta_{CB}$ | Cassie contact angle |
| $\Phi$ | Nonlinear function of $\theta_{CB}$ and $\theta_{Mdep}$ |

To analyze the surface characteristics related to equation B2, it is extremely important to convert the fractional exponents of $\varphi$ to linear formulations. To aid the simplification, the number B1 is rearranged as the product of 4 non-linear functions of $\cos\theta_{CB}$, such that two of the functions are reciprocals to each other (equation B3).

$$1 = (2 + \cos\theta_{CB})^{\frac{2}{3}}(2 + \cos\theta_{CB})^{-\frac{2}{3}}(1 - \cos\theta_{CB})^{\frac{1}{3}}(1 - \cos\theta_{CB})^{-\frac{1}{3}} \qquad \text{B3)}$$

The expression $\varphi$ is multiplied with equation B3 (equation B4).

$$\varphi(1) = \varphi(2 + \cos\theta_{CB})^{\frac{2}{3}}(2 + \cos\theta_{CB})^{-\frac{2}{3}}(1 - \cos\theta_{CB})^{\frac{1}{3}}(1 - \cos\theta_{CB})^{-\frac{1}{3}} \qquad \text{B4)}$$

On simplification, $\varphi$ is re-written as a product of two linear functions and two non-linear functions, where the nonlinear functions are expressed as ratios of $\cos\theta_{CB}$ (equation B5).

$$\varphi = (2 + \cos\theta_{CB})(1 - \cos\theta_{CB})\left(\frac{1-\cos\theta_{Mdep}}{1-\cos\theta_{CB}}\right)^{\frac{1}{3}}\left(\frac{2+\cos\theta_{Mdep}}{2+\cos\theta_{CB}}\right)^{\frac{2}{3}} \qquad \text{B5)}$$

Next, the part of the expression consisting of a nonlinear expression of $\cos\theta_{Mdep}$ must be linearized. The difference in the cosines of $\theta_{Mdep}$ and $\theta_{CB}$, $\delta$, plays an important role in the conversion of the nonlinear function to its linear counterpart (equation B6).

$$\delta = \cos\theta_{Mdep} - \cos\theta_{CB} \; ; \; \therefore \cos\theta_{Mdep} = \delta + \cos\theta_{CB} \qquad \text{B6)}$$

The two nonlinear functions present in $\varphi$ (equation B5) are individually simplified. The cosine of $\theta_{Mdep}$ is expressed in terms of $\delta$, and

$$\left(\frac{1-\cos\theta_{Mdep}}{1-\cos\theta_{CB}}\right)^{\frac{1}{3}} = \left(1 - \frac{\delta}{1-\cos\theta_{CB}}\right)^{\frac{1}{3}} \qquad \text{B7)}$$

$$\left(\frac{2+\cos\theta_{Mdep}}{2+\cos\theta_{CB}}\right)^{\frac{2}{3}} = \left(1 + \frac{\delta}{2+\cos\theta_{CB}}\right)^{\frac{2}{3}} \qquad \text{B8)}$$

In the next steps, binomonal equation of fractional exponents is used to simplify and expand $\varphi$. The binomial expansion of an algebraic function with a coefficient *s* and a fractional exponent *n* is given as follows.

$$(1+s)^n = 1 + ns + \frac{n(n-1)}{2!}s^2 + \frac{n(n-1)(n-2)}{3!}s^3 \\ + \frac{n(n-1)(n-2)(n-3)}{4!}s^4 \quad \text{B9)}$$

Using binomial expansion, equations B7 and B8 are simplified to the 3rd term (equations B10-B13).

$$\left(1 - \frac{\delta}{1-\cos\theta_{CB}}\right)^{\frac{1}{3}} = 1 + \frac{1}{3}\left(-\frac{\delta}{1-\cos\theta_{CB}}\right) + \frac{1}{3}\left(\frac{1}{3}-1\right)\frac{1}{2!}\left(-\frac{\delta}{1-\cos\theta_{CB}}\right)^2 \quad \text{B10)}$$

$$\left(1 - \frac{\delta}{1-\cos\theta_{CB}}\right)^{\frac{1}{3}} = 1 - \frac{\delta}{3(1-\cos\theta_{CB})} - \frac{\delta^2}{9(1-\cos\theta_{CB})^2} \quad \text{B11)}$$

$$\left(1 + \frac{\delta}{2+\cos\theta_{CB}}\right)^{\frac{2}{3}} = 1 + \frac{2}{3}\left(\frac{\delta}{2+\cos\theta_{CB}}\right) + \frac{2}{3}\left(\frac{2}{3}-1\right)\frac{1}{2!}\left(\frac{\delta}{2+\cos\theta_{CB}}\right)^2 \quad \text{B12)}$$

$$\left(1 + \frac{\delta}{2+\cos\theta_{CB}}\right)^{\frac{2}{3}} = 1 + \frac{2\delta}{3(2+\cos\theta_{CB})} - \frac{\delta^2}{9(2+\cos\theta_{CB})^2} \quad \text{B13)}$$

Upon simplification, equations B11 and B13 are multiplied. Since $\delta$ is the difference between two cosines, its absolute value is always less than unity. Hence, the coefficients of the higher exponents $\delta$ ($\delta^3$ and $\delta^4$) are neglected (equation B 14).

$$\left(1 - \frac{\delta}{1-\cos\theta_{CB}}\right)^{\frac{1}{3}}\left(1 + \frac{\delta}{2+\cos\theta_{CB}}\right)^{\frac{2}{3}} = 1 - \frac{\delta\cos\theta_{CB}}{(1-\cos\theta_{CB})(2+\cos\theta_{CB})} - \frac{\delta^2}{(1-\cos\theta_{CB})^2(2+\cos\theta_{CB})^2} \quad \text{B14)}$$

Equation 14 is substituted in equation B5 (equation B15).

$$\varphi = (2 + \cos\theta_{CB})(1 - \cos\theta_{CB})\left(1 - \frac{\delta\cos\theta_{CB}}{(1-\cos\theta_{CB})(2+\cos\theta_{CB})} \right. \\ \left. - \frac{\delta^2}{(1-\cos\theta_{CB})^2(2+\cos\theta_{CB})^2}\right) \quad \text{B15)}$$

$$\varphi = (2 + \cos\theta_{CB})(1 - \cos\theta_{CB}) - \delta\cos\theta_{CB} - \frac{\delta^2}{(1-\cos\theta_{CB})(2+\cos\theta_{CB})} \quad \text{B16)}$$

The parameter $\delta$ is expressed in terms of $\theta_{CB}$ and $\theta_{Mdep}$ (equation B17).

$$\varphi = 2 - \cos\theta_{CB}(1 + \cos\theta_{Mdep}) - \frac{(\cos\theta_{Mdep} - \cos\theta_{CB})^2}{(1-\cos\theta_{CB})(2+\cos\theta_{CB})} \qquad \text{B17)}$$

The simplified form of $\varphi$ is substituted to equation B2 (equation B18).

$$\frac{4ah\cos\theta_Y(1+\cos\theta_{Mdep})}{(a+b)^2} - \frac{(\cos\theta_{Mdep} - \cos\theta_{CB})^2}{(1-\cos\theta_{CB})(2+\cos\theta_{CB})} = 0 \qquad \text{B18)}$$

Equation B18 is re-arranged to generate a quadratic expression of $\cos\theta_{Mdep}$ (equation B19).

$$\cos^2\theta_{Mdep} + \cos\theta_{Mdep}(-2\cos\theta_{CB} - \tau) + (\cos^2\theta_{CB} - \tau) = 0 \qquad \text{B19)}$$

where $\tau = \dfrac{4ah\cos\theta_Y(1-\cos\theta_{CB})(2+\cos\theta_{CB})}{(a+b)^2}$

Equation B19 marks the first instance, where the APCA for a depinned state $\theta_{Mdep}$ is expressed as a function of $\theta_Y$, $h$, $a$, $b$. Thus, to have a realizable $\theta_{Mdep}$, the discriminant of equation B19 ($\Delta$) must be positive (necessary condition, equation B20). In addition, one root of equation B19 must possess a realizable value (sufficient condition, equation B1).

$$\Delta = (-2\cos\theta_{CB} - \tau)^2 - 4(\cos^2\theta_{CB} - \tau) = \tau(\tau + 4 + 4\cos\theta_{CB}) \qquad \text{B20)}$$

NECESSARY CONDITION: $\Delta > 0$

The discriminant ($\Delta$) is expressed as the product of two functions, namely $\tau$ and $(\tau+4+4\cos\theta_{CB})$. For $\Delta > 0$, both the functions must possess the identical sign. A case study is performed, where we analyze the ramifications when both the functions are positive (case i) or negative (case ii).

Case i: $\tau > 0$, and $(\tau + 4 + 4\cos\theta_{CB}) > 0$

Case ii: $\tau < 0$ and $(\tau + 4 + 4\cos\theta_{CB}) < 0$

The above mentioned cases are analyzed as follows.

Case i

The function $\tau$ is a product of several expressions.

$$\tau = \frac{4ah\cos\theta_Y(1-\cos\theta_{CB})(2+\cos\theta_{CB})}{(a+b)^2} > 0 \qquad \text{B21)}$$

The terms $a$, $(a+b)^2$, $h$, $(2+\cos\theta_{CB})$ and $(1-\cos\theta_{CB})$ are each positive. Thus, the expression is true when $\cos\theta_Y > 0$.

$$\cos\theta_Y > 0; 0° < \theta_Y < 90° \qquad \text{B22)}$$

The domain of case i is mutually exclusive to that in the current discussion ($\theta_Y > 90°$, equation B 1). Since case i falls beyond the scope of this discussion, it is not analyzed any further.

Case ii

For case ii to be true, each of the expressions $\tau$ and $(\tau+4+4\cos\theta_{CB})$ ought to be negative. From the analysis of case i, it can be inferred that $\theta_Y > 90°$ (which is compatible with the domain of $\theta_Y$ in discussion) is associated with $\tau < 0$. Thus, the sufficient condition can be determined by pinpointing the surface characteristics with $\tau+4+4\cos\theta_{CB} < 0$ (equation B23).

$$\frac{4ah\cos\theta_Y(1-\cos\theta_{CB})(2+\cos\theta_{CB})}{(a+b)^2} + 4 + 4\cos\theta_{CB} < 0 \qquad \text{B23)}$$

Equation B23 is simplified to render the surface characteristics for case ii, and hence, the phenomenon of a depinned state for a surface with $\theta_Y > 90°$. Since $\cos\theta_Y < 0$, $|\cos\theta_Y| = -\cos\theta_Y$. Equation B23 is re-arranged to give a minimum permissible value for penetration depth $h$ (equation B25).

$$\frac{4ah|\cos\theta_Y|(1-\cos\theta_{CB})(2+\cos\theta_{CB})}{(a+b)^2} > 4 + 4\cos\theta_{CB} \qquad \text{B24)}$$

$$h > \frac{(a+b)^2}{a} \frac{(1+\cos\theta_{CB})}{(1-\cos\theta_{CB})(2+\cos\theta_{CB})|\cos\theta_Y|} \qquad \text{B25)}$$

Thus, to have $\Delta > 0$, the penetration depth has a minimum value determined by $a$, $b$, $\theta_Y$. It is seen that $h$ typically assumes values of the order of mm, much higher than the μm sized pillar height *c*. This clearly shows that in general, it is not feasible to have a penetration with $\theta_Y > 90°$. In the following section, the sufficient condition to have a mathematically deductible $\theta_M$ is described.

## Sufficient condition to have a $\theta_{Mdep}$ with $\theta_Y > 90°$

Since the general equation of wettability has been simplified to a quadratic equation of $\cos \theta_M$ (equation B19), feasible results can be obtained when $-1 < \cos \theta_{Mdep} < 1$. The sufficient condition is analyzed for the case $\theta_Y > 90°$. The simplified form of the general equation of wettability is expressed in the form of a quadratic equation (equation B26).

$$\cos^2 \theta_{Mdep} + \beta \cos \theta_{Mdep} + \chi = 0 \qquad \text{B26)}$$

Where $\beta = -\tau - 2 \cos \theta_{CB}$ ; $\chi = -\tau + \cos^2 \theta_{CB}$

So, to have a valid $\theta_{Mdep}$, $-1 < \cos \theta_{Mdep} < 1$

The root corresponding to $\cos \theta_{Mdep} = \frac{-\beta - \sqrt{\beta^2 - 4\chi}}{2}$ is ignored as it renders values less than -1, the minimum possible value of $\cos \theta_{Mdep}$. The other root, namely $\cos \theta_{Mdep} = \frac{-\beta + \sqrt{\beta^2 - 4\chi}}{2}$ is considered, and substituted to equation B1 (equation B27).

$$-1 \leq \cos \theta_{Mdep} = \frac{-\beta + \sqrt{\beta^2 - 4\chi}}{2} \leq 1 \qquad \text{B27)}$$

On simplification, equation B27 gives rise to an inequality (equation B28). Now, both the inherent inequalities comprising equation B28 must be correct.

$$\beta - 2 \leq \sqrt{\beta^2 - 4\chi} \leq \beta + 2 \qquad \text{B28)}$$

To further analyze the result, the inequality must be squared. It should be noted that the inequality, on being squared, may not necessarily retain its sign. The modulus of each term must be squared and compared. To demonstrate this, a corollary is presented as follows.

## Corollary

On squaring the inequality $-4 < 2 < 5$ without changing signs, a wrong result is obtained, i.e. $16 < 4 < 25$. The squared inequality is not correct, since $16 > 4$. The domain of $\beta$ plays a very crucial role in further analysis.

$$\forall \theta_Y > 90°; \because \tau < 0; \beta > 0; \therefore |\beta + 2| > |\beta - 2| \qquad \text{B29)}$$

For $\theta_Y > 90°$, the inequality can be simply squared without changing signs.

$$\forall \theta_Y > 90°; \because \tau < 0; \beta > 0; \therefore |\beta + 2| > |\beta - 2| \qquad \text{B30)}$$

$$|\beta - 2| \leq \sqrt{\beta^2 - 4\chi} \leq |\beta + 2| \qquad \text{B31)}$$

$$(\beta - 2)^2 \leq \beta^2 - 4\chi \leq (\beta + 2)^2 \qquad \text{B32)}$$

Inequality B32 is simplified to generate inequality B33.

$$-1 - \beta \leq \chi \leq \beta - 1 \qquad \text{B33)}$$

Substituting $\beta = -\tau - 2\cos\theta_{CB}$; $\chi = -\tau + \cos^2\theta_{CB}$, inequality B33 is simplified in the following steps to render inequality B36.

$$-1 + \tau + 2\cos\theta_{CB} \leq -\tau + \cos^2\theta_{CB} \leq -\tau - 2\cos\theta_{CB} - 1 \qquad \text{B34)}$$

$$-1 + 2\tau + 2\cos\theta_{CB} - \cos^2\theta_{CB} \leq 0 \leq -2\cos\theta_{CB} - 1 - \cos^2\theta_{CB} \qquad \text{B35)}$$

$$2\tau - (1 - \cos\theta_{CB})^2 \leq 0 \leq -(1 + \cos\theta_{CB})^2 \qquad \text{B36)}$$

The above inequality suggests that $0 \leq -(1 + \cos\theta_{CB})^2$, which is absurd. Hence, it can be inferred that no sufficient condition exists for a depinned state with $\theta_Y > 90°$. Since neither the necessary condition, nor the sufficient condition render mathematically plausible surface characteristics, it is found that surface energy minimization cannot solely account for a depinned state for surfaces with $\theta_Y > 90°$.

**Supporting information C: Proof of the existence of a intermediate wetting state**

The evolution of the apparent contact angle with increasing pillar spacing to pillar width ratios follows a unique trend for surfaces with $\theta_Y > 90°$ {He, 2003 #127;Zhu, 2006 #1313;Zhang, 2007 #1312;Barbieri, 2007 #1310;Varanasi, 2009 #893}. In the following figure, the APCA is plotted against spacing to width ratio for square pillar geometry with pillar width of 25 μm and Young's contact angle ($\theta_Y$) of 114° {He, 2003 #127}. Cassie and Wenzel equations are also plotted for the same series of surfaces (Figure C1).

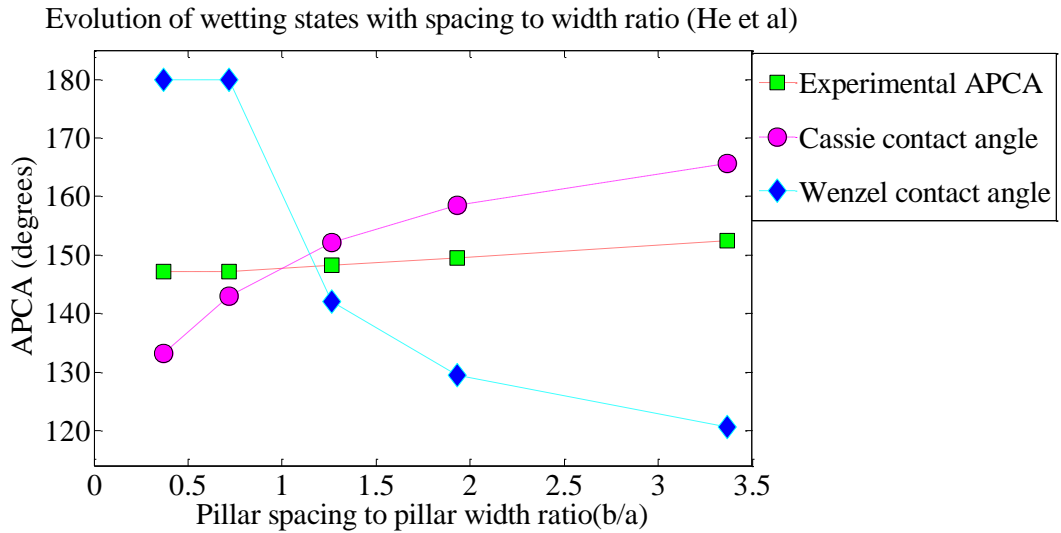

Figure C1 Variation of APCA with spacing to width ratio

There exists a unique spacing to width ratio, also known as critical spacing to width ratio ($b/a$=1.15) for a given surface chemistry for which the calculated Cassie and Wenzel contact angles are equal. For b/a ratios greater than the critical b/a ratio, the Wenzel state becomes energetically favorable to the Cassie state, and the static contact angle assumes values in between those predicted by Cassie and Wenzel models. Here, for b/a ratios between 1.15 and 3.0, the APCA is closer to the Cassie contact angle. Light transmission experiments revealed the existence of the liquid-air interface under the apex of the surface roughness for b/a ratios exceeding 1.15. {He, 2003 #127;Varanasi, 2009 #893}. For very high b/a ratios (> 3), the static contact angle measurements follow the Wenzel model. Erbil et al. have investigated the existing static contact angle measurements for a set of surfaces with square pillar topologies, distinct chemistries, and increasing spacing to width ratios {Erbil, 2009 #1209;Zhang, 2007 #1312;Zhu, 2006 #1313}. To understand the wetting states of the aforementioned set of data, the collected APCAs from the above experiments were listed. Assuming the wetting states were either Cassie (equation C1) or Wenzel, and with the knowledge of the YCA of the surfaces used ($\theta_Y$), the experimentally determined APCAs ($\theta_{exp}$) were substituted into the Cassie equation (equation C 2). The solid fraction $f_{exp,Erbil}$, as obtained from the substitution was expressed in terms of $\theta_{exp}$ and $\theta_Y$ (equation C3), and compared to the solid fraction as defined by the surface geometry ($f$).

$$\cos \theta_{CB} = f \cos \theta_Y + f - 1 \qquad \text{C1)}$$

$$\cos \theta_{exp} = f_{exp,Erbil} \cos \theta_Y + f_{exp,Erbil} - 1 \qquad \text{C2)}$$

$$f_{exp,Erbil} = \frac{(1+\cos \theta_{exp})}{(1+\cos \theta_Y)} \qquad \text{C3)}$$

The change in solid fraction, measured as $\Delta f_{exp,Erbil}$ (equation C4), is tabulated (table C1).

$$\Delta f_{exp,Erbil} = f - f_{exp,Erbil} = \frac{(\cos \theta_{CB} - \cos \theta_{exp})}{(1+\cos \theta_Y)} \qquad \text{C4)}$$

A negative change, i.e. $\Delta f_{exp,Erbil} < 0$ denotes penetration of water in the roughness valleys. It has been seen that for the 31 surfaces investigated, only 6 surfaces exhibit a negative change. For the remaining 25 cases, a positive change is recorded. This means that the liquid does not completely wet the apex of the roughness features, which is absurd. The error in the determination of solid fraction arises from overestimation of the contribution of the liquid-air fraction in equation 1 {Milne, 2012 #1315}. We postulate that the area occupied by the liquid-air interface (liquid-air fraction) is independent of the degree of liquid penetration inside the roughness valleys (solid fraction). Equation C2 is corrected (equation C5), and the corrected solid fraction $f_{exp,corrected}$, is determined (equation C6).

$$\cos \theta_{exp} = f_{exp,corrected} \cos \theta_Y + f - 1 \qquad \text{C5)}$$

The change in solid fraction ($\Delta f_{exp,corrected}$) is calculated (equations C6 and C7) and listed for the set of 31 surfaces (table C1).

$$f_{exp,corrected} = \frac{(\cos\theta_{exp}+1-f)}{\cos\theta_Y} \qquad \text{C6)}$$

$$\Delta f_{exp,corrected} = f - f_{exp,corrected} = \frac{(\cos\theta_{CB}-\cos\theta_{exp})}{\cos\theta_Y} \qquad \text{C7)}$$

Only 6 of the 31 surfaces show a positive change in solid fraction (highlighted in grey). The remaining 25 surfaces exhibit a negative change in solid fraction, thereby indicating a penetration in the roughness valleys. Thus, it is safe to infer that the intermediate state exists for surfaces with $\theta_Y > 90°$.

Table C1 Evidence of an intermediate state: penetration observed for 25 of 28 surfaces.

|  | *Surface* | $\theta_Y$ (°) | *F* | *Δf*$_{CB,Erbil}$ | *Δf*$_{CB,corrected}$ |
|---|---|---|---|---|---|
|  | 1. |  | 0.24 | 0.05 | -0.12 |
|  | 2. |  | 0.29 | 0.07 | -0.17 |
|  | 3. |  | 0.41 | 0.17 | -0.41 |
|  | 4. |  | 0.45 | 0.22 | -0.53 |
|  | 5. |  | 0.50 | 0.27 | -0.65 |
|  | 6. |  | 0.59 | 0.33 | -0.80 |
|  | 7. |  | 0.77 | 0.49 | -1.18 |
|  | 8. |  | 0.97 | 0.04 | -0.10 |
| {Zhang, 2007 #1312} | 9. | 107 | 0.14 | -0.06 | 0.15 |
|  | 10. |  | 0.29 | 0.09 | -0.21 |
|  | 11. |  | 0.40 | 0.18 | -0.43 |
|  | 12. |  | 0.45 | 0.20 | -0.49 |
|  | 13. |  | 0.47 | 0.22 | -0.53 |
|  | 14. |  | 0.60 | 0.34 | -0.82 |
|  | 15. |  | 0.70 | 0.41 | -0.99 |
|  | 16. |  | 0.94 | 0.12 | -0.29 |
|  | 17. |  | 0.97 | -0.03 | 0.07 |
|  | 18. |  | 0.21 | 0.03 | -0.05 |
|  | 19. |  | 0.32 | 0.15 | -0.27 |
|  | 20. |  | 0.38 | 0.15 | -0.26 |
|  | 21. |  | 0.44 | 0.17 | -0.30 |
|  | 22. |  | 0.47 | 0.25 | -0.46 |
| {Zhu, 2006 #1313} | 23. | 111 | 0.72 | 0.38 | -0.68 |
|  | 24. |  | 0.39 | 0.14 | -0.26 |
|  | 25. |  | 0.33 | 0.09 | -0.16 |
|  | 26. |  | 0.29 | 0.05 | -0.10 |
|  | 27. |  | 0.21 | 0.03 | -0.05 |
|  | 28. |  | 0.14 | -0.02 | 0.04 |

**Supporting information D: Determination of quasi-static limit for robustness**

The antiwetting pressure must be higher than 117.23 Pa for a quasi-statically robust surface. Equation 17 of the main text (shown here as equation D1) is solved, where all the parameters constituting the antiwetting pressure are converted to their respective SI units. The pillar width and spacing, originally expressed in µm are converted to m. The expression is simplified (equations D1- D6) to generate the quasi-static limit of spacing to width ratios.

$$P_{antiwetting} = -\frac{4\gamma_{LA} a \cos\theta_Y}{b(2a+b)} = \frac{4\gamma_{LA} a |\cos\theta_Y|}{b(2a+b)} \quad \text{17.}$$

$$\frac{4 \times 0.072 \times a |\cos\theta_Y| 10^{-6}}{b(2a+b) 10^{-12}} > 117.23 \; Nm^{-2} \quad \text{D1)}$$

$$\frac{a|\cos\theta_Y|}{b(2a+b)} 10^6 > 407.06 \; Nm^{-2} \quad \text{D2)}$$

$$\frac{|\cos\theta_Y|}{a} \frac{10^6}{(1+\frac{b}{a})^2 - 1} > 407.06 \; Nm^{-2} \quad \text{D3)}$$

$$(1+\frac{b}{a})^2 - 1 \leq \frac{10^6 |\cos\theta_Y|}{407.06 a} \quad \text{D4)}$$

$$\frac{b}{a} \leq \sqrt{1 + \frac{2456.64 |\cos\theta_Y|}{a}} - 1 \quad \text{D5)}$$

$$\frac{b}{a} \leq \left(\frac{b}{a}\right)_{QS} = \sqrt{1 - \frac{2456.64 \cos\theta_Y}{a}} - 1 \quad \text{D6)}$$

In order to have a quasi-static limit, the quasi-static spacing to width ratio must exceed its critical counterpart (equation D7). Expressions for both the limits are substituted, and the inequality is simplified to determine the domain of *a*, *b* and $\theta_Y$ (equations D7- D14). A unique relationship is established between the height to width ratio and the surface chemistry (equation D15).

$$\left(\frac{b}{a}\right)_{critical} \leq \left(\frac{b}{a}\right)_{QS} \quad \text{D7)}$$

$$\sqrt{1 - \frac{4c \cos\theta_Y}{a(1+\cos\theta_Y)}} - 1 \leq \sqrt{1 - \frac{2456.64 \cos\theta_Y}{a}} - 1 \quad \text{D8)}$$

$$\sqrt{1 + \frac{4c|\cos\theta_Y|}{a(1-|\cos\theta_Y|)}} - 1 \leq \sqrt{1 + \frac{2456.64 |\cos\theta_Y|}{a}} - 1 \quad \text{D9)}$$

$$\frac{4c|\cos\theta_Y|}{a(1-|\cos\theta_Y|)} \leq \frac{2456.64|\cos\theta_Y|}{a} \qquad \text{D10)}$$

$$\frac{c}{(1-|\cos\theta_Y|)} \leq 614.16 \qquad \text{D11)}$$

$$1 - |\cos\theta_Y| \geq \frac{c}{614.16} \qquad \text{D12)}$$

$$1 + \cos\theta_Y \geq \frac{c}{614.16} \qquad \text{D13)}$$

$$\cos\theta_Y \geq \frac{c}{614.16} - 1 \qquad \text{D14)}$$

$$\theta_Y \leq \cos^{-1}(\frac{c}{614.16} - 1) \qquad \text{D15)}$$